\documentclass[twoside,a4]{ilcws08}
\usepackage[latin1]{inputenc}
\usepackage[dvips]{graphicx,epsfig,color}
\usepackage{wrapfig,rotating}
\usepackage{amssymb,amsmath,array}

\begin{document}
\title{
Detector Optimization for SiD using PFA} 
\author{Marcel Stanitzki
\vspace{.3cm}\\
STFC-Rutherford Appleton Laboratory\\
Chilton, Didcot Oxfordshire OX11 0QX - United Kingdom
}

\maketitle

\begin{abstract}
A summary of the optimization of the SiD detector is given. To optimize its
performance in terms of Particle Flow Algorithms (PFA) , five basic detector
parameters have been varied and the impact on the obtained energy resolution
using Particle Flow Algorithms has been studied using di-jets events. Finally
the optimized detector used for the Letter of Intent (LoI) is briefly summarized
as a result from these studies.\end{abstract}

\section{Introduction}
The SiD detector\cite{sid01} concept for the ILC\footnote{International Linear
Collider} has been designed with the Particle Flow Algorithm approach in mind. The PFA approach actually drives a lot of the design
choices such as locating the highly granular calorimetry within the
superconducting coil. In order to maximize the performance of the SiD detector
concept, events were simulated  using several different detector variants and these events
were run through the full PFA chain. This allowed us to choose an optimal
configuration for PFA while keeping other constraints like engineering or costs
in mind. The quantity being optimized is the Jet Energy Resolution, which
maximizes the physics potential of the SiD detector. 

The SiD detector (sid01) consists of a 5-layer Vertex Detector using Silicon pixels, a 5
layer tracker using silicon strip detector, a 30 layer electromagnetic
calorimeter (ECAL) using Silicon-Tungsten, a 34 layer hadronic calorimeter
(HCAL) using Iron and RPC's\footnote{Resistive Plate Chambers}. Outside of the
HCAL is the 5 T solenoid and the Muon system integrated in the iron flux return
yoke.

\section{The Software Setup}
\begin{wraptable}{l}{0.5\columnwidth}
\centerline{
\begin{tabular}{|l|r|r|}\hline
		 &LDC00Sc	&SiDish	\\\hline
ECAL inner radius&       1.7 m   &1.25 m \\\hline
ECAL length	 &       2.7 m   &1.7 m \\\hline
ECAL layers	 &       30+10   &20+10 \\\hline
ECAL material	 &       SiW     &SiW \\\hline
HCAL layers	 &       40      &40 \\\hline
HCAL material	 &       Fe-Scint&Fe-Scint\\\hline
B Field		 &       4 T     &5 T\\ \hline
\end{tabular}
}\caption{The parameters of the {\tt LDC00Sc} and {\tt SiDish}  detector models.}
\label{tab:sidish}
\end{wraptable}
The PFA algorithm chosen for this study is PandoraPFA 2.01\cite{Thomson:2008zz}, which has been developed
by Mark Thomson. This requires the use of the {\tt Mokka}
package\cite{ MoradeFreitas:2004sq} for the {\tt GEANT4}\cite{Agostinelli:2002hh} based
detector simulation and  {\tt Marlin}\cite{Gaede:2006pj} for event reconstruction and not SiD's
tool chain of  {\tt SLIC}\cite{Graf:2006ei} for the detector simulation and {\tt org.lcsim} for the
event reconstruction. As there is currently no accurate SiD  detector simulation
available within {\tt Mokka}, a SiD-look-alike, the {\tt SiDish} has been derived
from the {\tt LDC00Sc} model available in {\tt Mokka}. The {\tt LDC00Sc} model is using the old TESLA detector design \cite{Behnke:2001qq}. 
As the calorimetry is the largest cost driver of any PFA-based ILC detector, a
detector which such a deep and long calorimeter is considered to be too
expensive to be an realistic option. E.g. scaling up the ECAL increases its volume by a factor of $\approx$ 2.7.

Since the {\tt LDC00Sc} model has a TPC\footnote{Time Projection Chamber} and there was also no all-silicon tracking software available at the time, the TrackCheater
from the MarlinReco package was used. For the {\tt GEANT4} physics list, we used the LCPHYS
list, which has been recommended by WWS Software panel. The parameters of the simulated {\tt SiDish} detector are shown in Tab. \ref{tab:sidish}.

\begin{wrapfigure}{r}{0.72\columnwidth}
\centerline{
{\includegraphics[width=0.35\columnwidth]{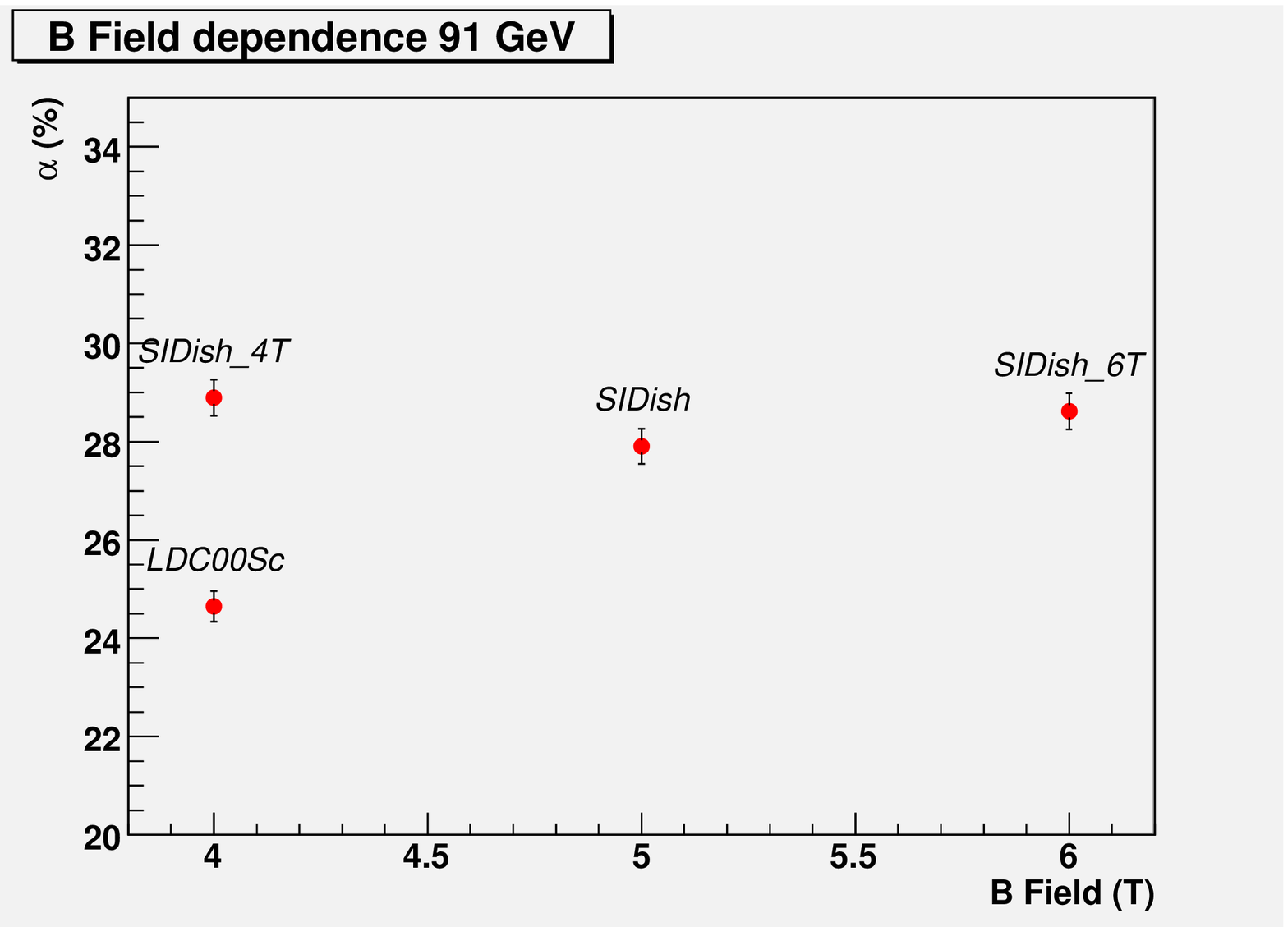}}
{\includegraphics[width=0.35\columnwidth]{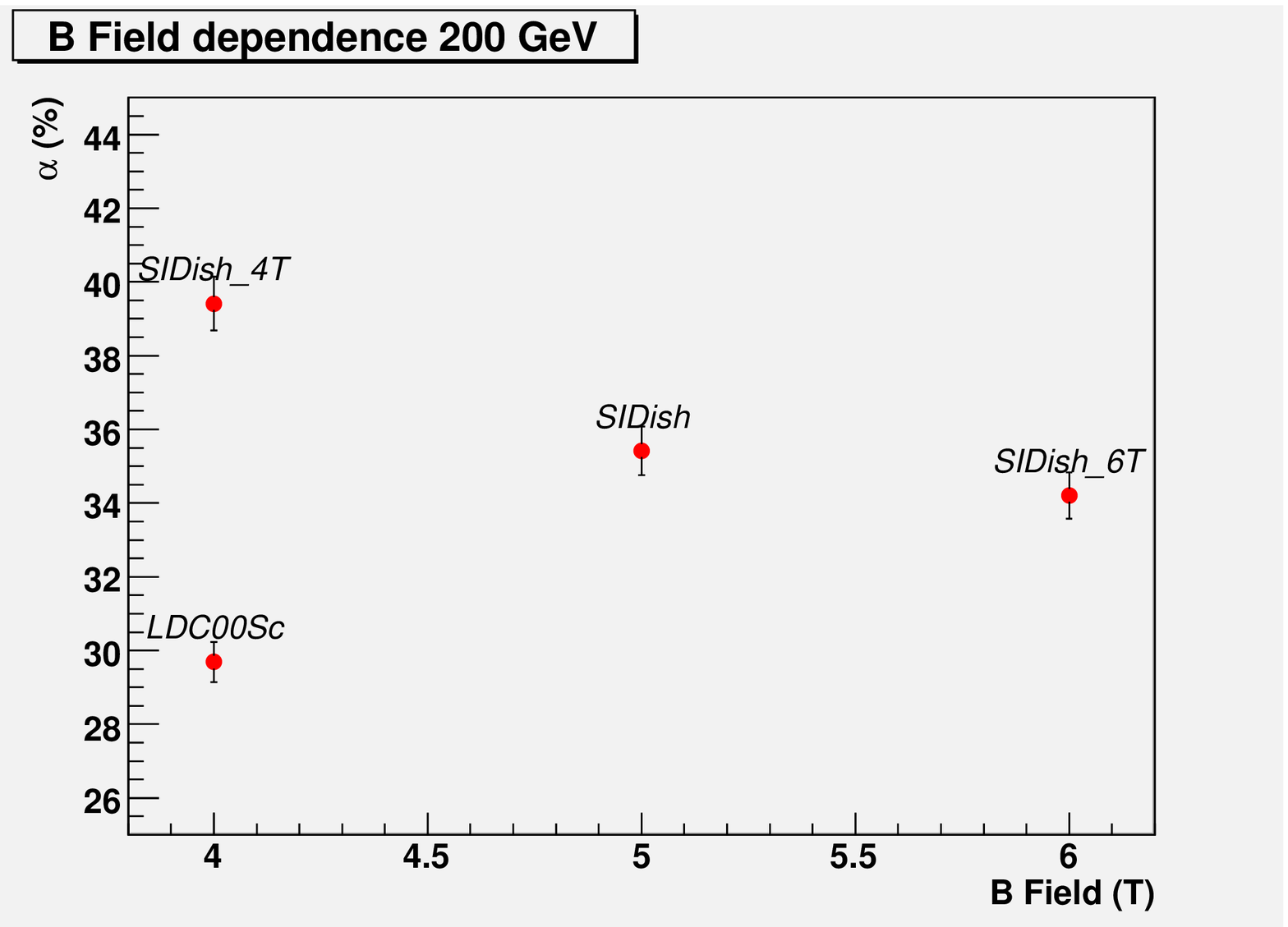}}
}
\caption{The dependence of the PFA on the B field for $Z \rightarrow q\bar{q} \,
q=u,d,s$  at $\sqrt{s}$=91 (left) and 200 GeV (right). The point obtained for {\tt LDC00Sc} is
shown for reference as well. }\label{Fig:bdep}
\end{wrapfigure}

It should be emphasized, that there are still fundamental differences in e.g.
the HCAL, as SiD uses a digital HCAL with RPC's for the readout instead of the scintillator readout in {\tt SiDish}.
Also there are differences in the general calorimeter layout and the tracker
material distribution. So one should only consider {\tt SiDish} as an
approximation of SiD in the framework of this study.

\section{Results}
The coordinate system of SiD follows the general coordinate system of many HEP
detectors, with the z-axis set to be along the beam axis. In this study, five main
detector parameters have been studied:
\begin{enumerate}
\item Magnitude of the B Field
\item ECAL inner Radius
\item ECAL inner length in z
\item HCAL depth in $\Lambda_{Iron}$
\item HCAL longitudinal segmentation
\end{enumerate}
In all studies we have used $Z \rightarrow q\bar{q} \,\, q=u,d,s$ events at
various energies unless mentioned otherwise. To measure the performance we use
$\alpha$ (in percent) as defined in 
\begin{equation}
\frac{\sigma_E}{E}=\frac{\alpha}{\sqrt{E}} \,\,\, \mathrm{with} \, \,\cos(\theta_{Thrust})<0.7 
\end{equation}
and as used by Thomson\cite{Thomson:2008zz}.
SiD has a 5 T field as baseline since the high field it is beneficial for
tracking and vertexing and helps suppressing the beam background. 

\subsection{The B field magnitude}
To study the impact of the B-Field, several versions of {\tt SiDish} have been
simulated with a B field from 4-6 T. The impact of the B field on the PFA performance is shown in Fig. \ref{Fig:bdep}. 
For a compact detector design like SiD, a 5 T field is to be preferred over a 4 T
field from the point of view of both PFA and tracking performance\cite{sid01}. 

\subsection{The inner radius of the ECAL}

\begin{wrapfigure}{r}{0.36\columnwidth}\centerline
{\includegraphics[width=0.36\columnwidth]{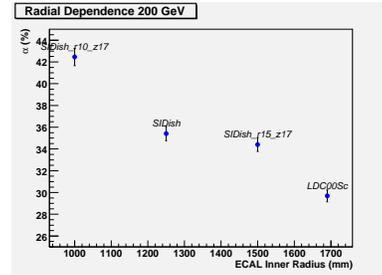}}
\caption{The dependence of the PFA on the inner radius of the ECAL for $Z \rightarrow q\bar{q} \,
q=u,d,s$  at $\sqrt{s}$=200 GeV. The point for obtained {\tt LDC00Sc} is
shown for reference as well.}\label{Fig:rdep}
\end{wrapfigure}

Due to the choice of a 5 T B field, the space inside the coil is limited due to
material and engineering constraints, leading
to a maximum inner radius of the ECAL of about 1.5 m. For this study the inner radius
of the ECAL was varied between 1.0 and 1.5 m using both events at $\sqrt{s}$=91
and 200 GeV (see Fig. \ref{Fig:rdep}). For a 5 T field, increasing the radius to
1.5 m only leads to a small gain, however decreasing the inner radius of the ECAL
to 1.0 m has a quite sizeable impact on the PFA performance.

\subsection{The inner length of the ECAL}
\begin{wrapfigure}{l}{0.72\textwidth}
{\includegraphics[width=0.36\textwidth]{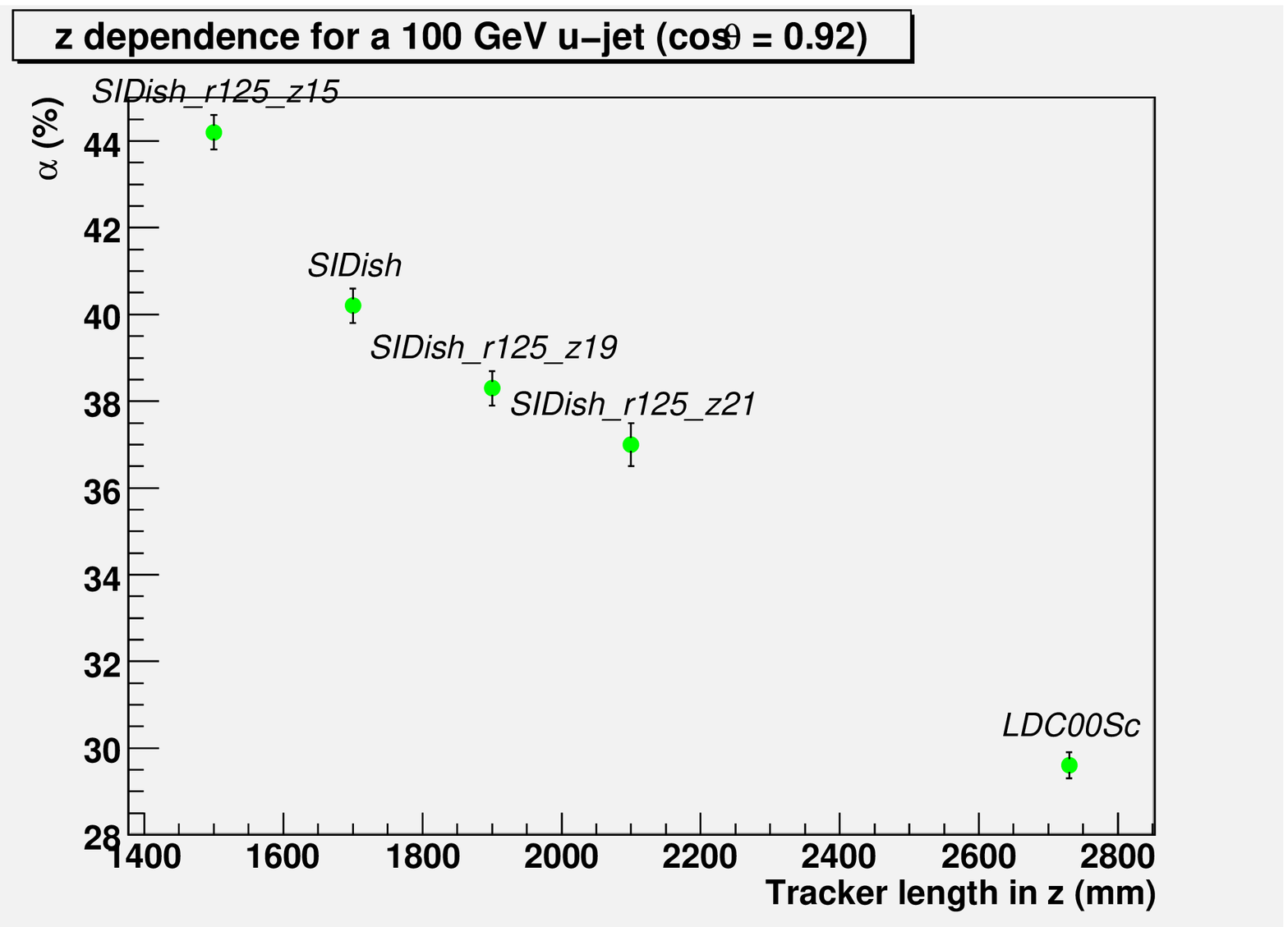}}
{\includegraphics[width=0.36\textwidth]{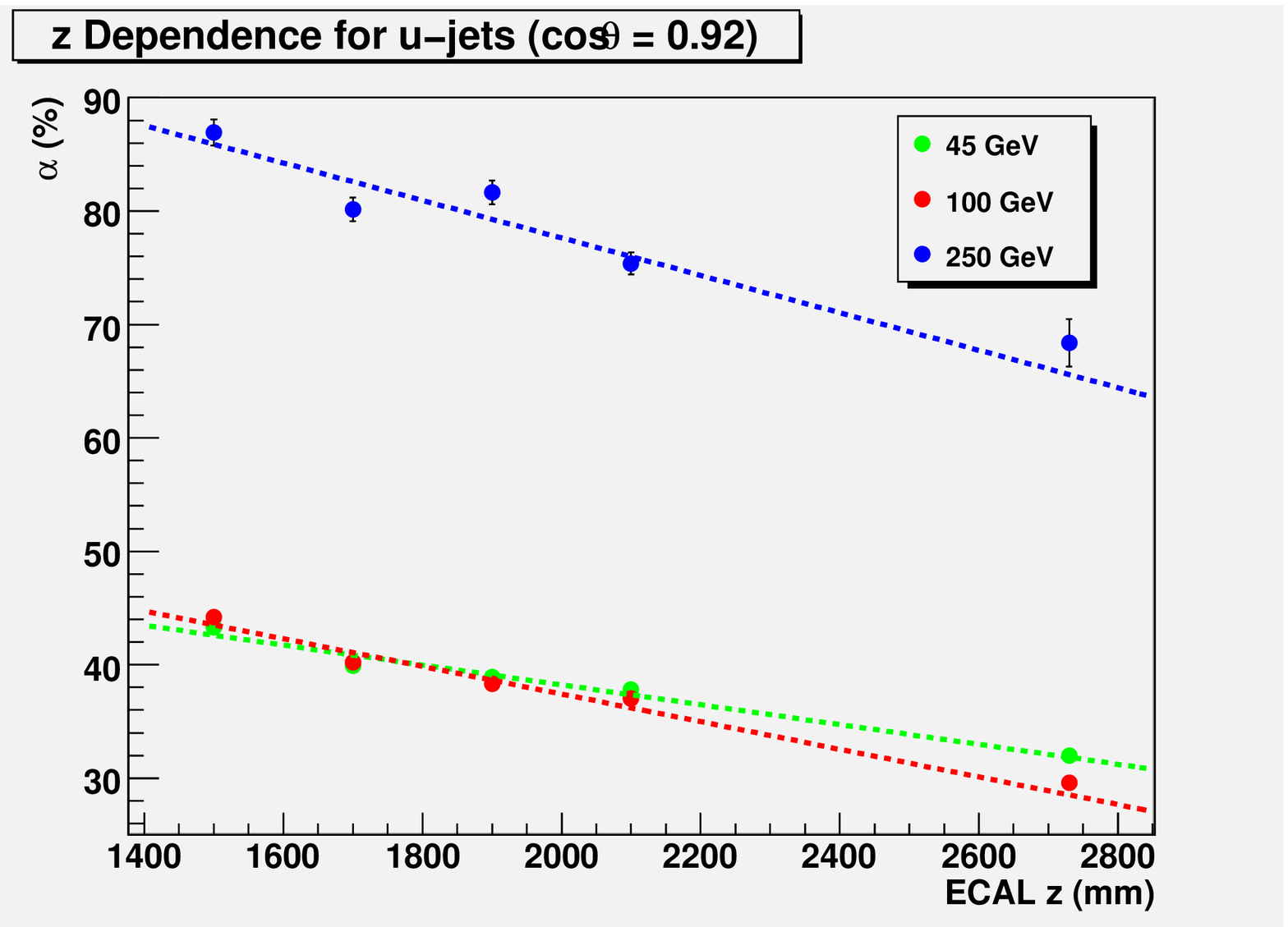}}
\caption{The z dependence for a single u-quark shown in left and the z
dependence for different energies (right). The point at 2.7 m is the
{\tt LDC00Sc} model result shown for comparison}\label{Fig:zdep}
\end{wrapfigure}

In this study a single u-quark with an energy of 45, 100 or 250 GeV was fired at
a fixed angle of $\cos(\theta)$=0.92 and then fragmented. This allows the study
the impact of inner z of the ECAL or in other words the impact of the length of
the tracker. The results from this study are shown in Fig. \ref{Fig:zdep}. It
can be clearly seen that a longer detector is beneficial for PFA.

\subsection{The optimization of the HCAL}
\begin{wrapfigure}{r}{0.74\columnwidth}
{\includegraphics[width=0.36\columnwidth]{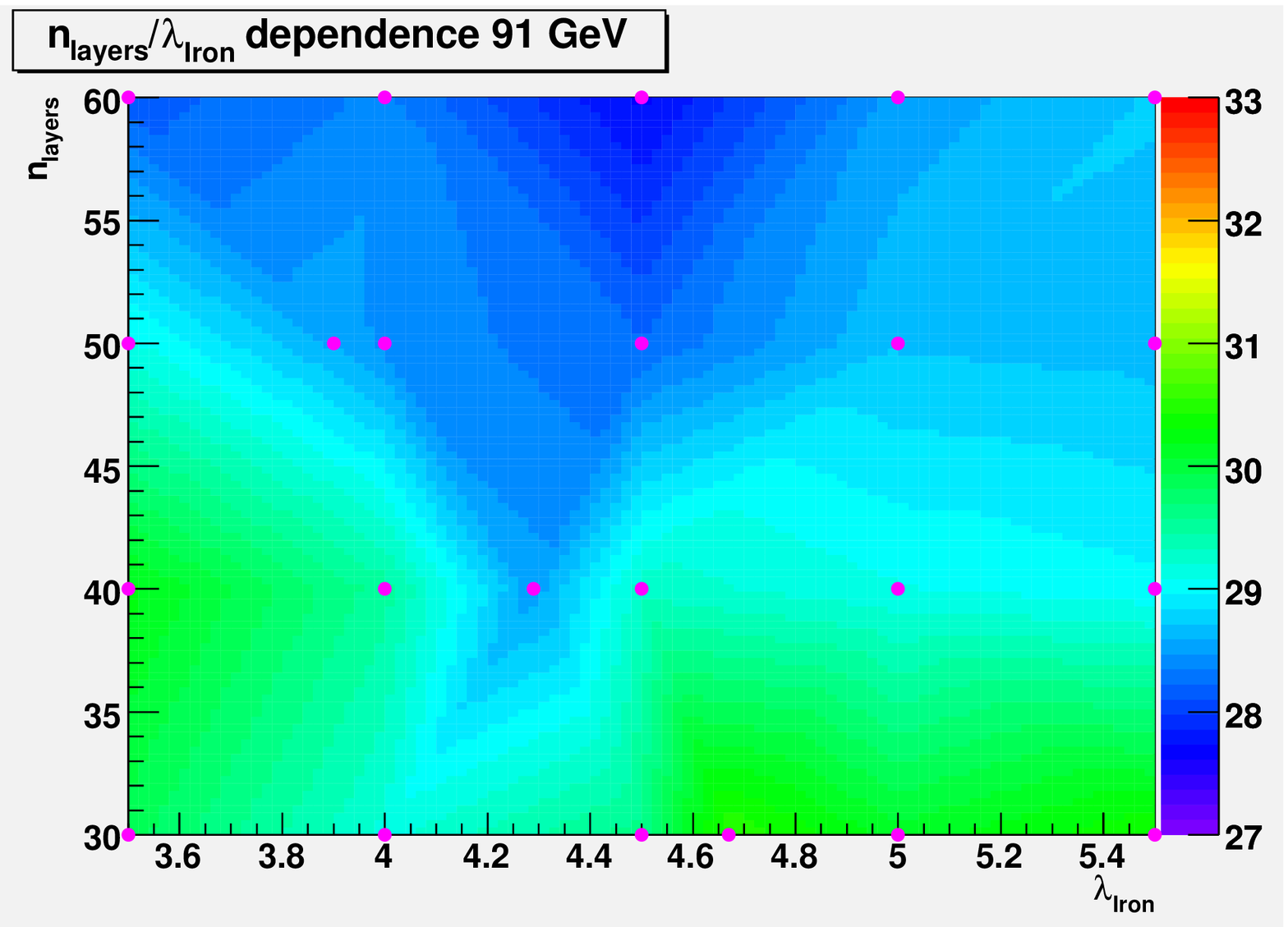}}
{\includegraphics[width=0.36\columnwidth]{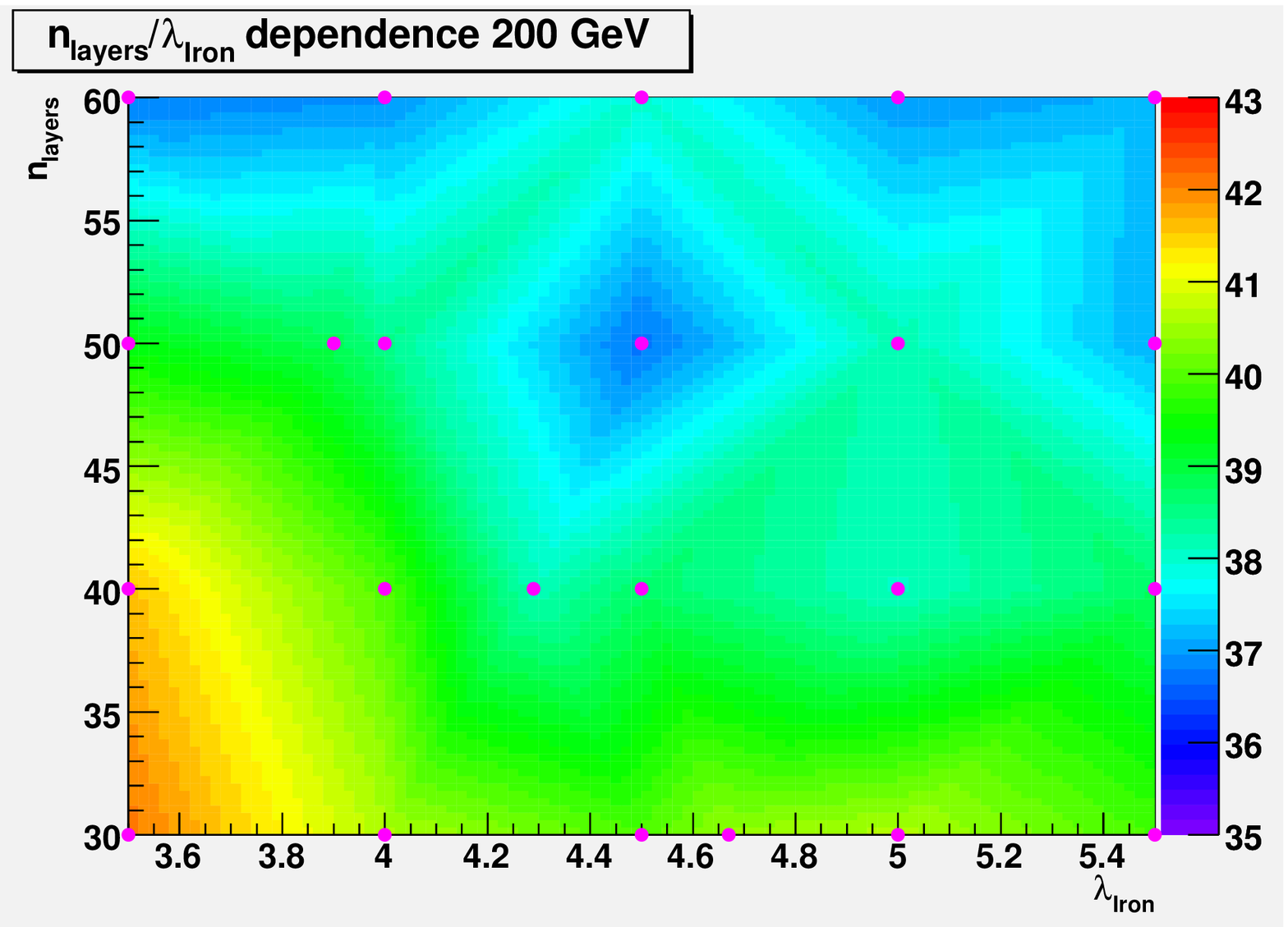}}
\caption{The dependence of the PFA on the HCAL depth and longitudinal
segmentation for $Z \rightarrow q\bar{q} \,
q=u,d,s$  at $\sqrt{s}$=91 (left) and 200 GeV (right). The pink dots indicate the
simulated points.}\label{Fig:hcal}
\end{wrapfigure}

For the HCAL, the depth in terms of $\Lambda_{Iron}$ and the number of layers
were both simultaneously optimized leading to about 36 different detector
configurations that have been evaluated, where both the thickness of the iron
absorber was varied between 3.5 and 5.5 $\Lambda_{Iron}$ and the number of
readout layers was varied between 30 and 60 layers. The detector baseline
(sid01) only had 34 HCAL layers with 4.0 $\Lambda_{Iron}$. From Fig. \ref{Fig:hcal} it is clear that
a deeper HCAL is beneficial for higher energies, however keeping engineering
constraints in mind, a 40 layer HCAL with 4.5 $\Lambda_{Iron}$ is an optimum for
SiD.

\section{The optimized SiD detector}
All the results shown before and input from other studies\cite{Breidenbach}
lead to two optimized detector variants, {\tt sid02} and {\tt sid02-stretch}.
The parameters of both are shown in Tab. \ref{tab:sidvar}. The variant with the
longer barrel has superior performance in the forward region which ranges from
2.5 - 4 \% (absolute) in the forward region, but due to engineering concerns,
SiD conservatively chose {\tt sid02} as the detector to use for the LoI. 

\begin{wraptable}{l}{0.65\columnwidth}
\centerline{
\begin{tabular}{|l|r|r|r|}\hline
				&sid01	&sid02&sid02-stretch  \\\hline
ECAL inner radius (m)		&1.25	&1.25 &1.25           \\\hline
ECAL inner Z (m)		&1.7	&1.7  &2.1            \\\hline
HCAL depth ($\Lambda_{Iron}$)	&4	&4.5  &4.5            \\\hline
HCAL layers			&34	&40   &40             \\\hline
B Field	(T)			&5	&5    &5              \\\hline
\end{tabular}
}\caption{The detector parameters for the sid01, sid02 and sid02-stretch models}
\label{tab:sidvar}
\end{wraptable}

\section{Conclusions and Outlook}
SiD has converged on an optimized detector for the LoI however the LoI will not be the
end of these studies. As SiD now has also a PFA algorithm\cite{Charles:2009ta}
in the {\tt org.lcsim} framework, a lot of these studies will continue using
SiD's own simulation and reconstruction framework. There will be continuous
work in exploring stretched detector designs with improved forward performance.
The current detector versions captures the current knowledge and lays out a
route for the post-LoI phase of SiD.

\section{Acknowledgments}
I would like to thank the SiD-PFA group for all their contributions and M.
Breidenbach, J.Jaros, M. Thomson, A. White and H. Weerts for useful discussions
on optimizing SiD.


\begin{footnotesize}

\end{footnotesize}

\end{document}